\documentclass[10pt]{article}
\usepackage[top=1in,bottom=1in,left=1.25in,right=1.25in]{geometry}
\usepackage{txfonts}
\usepackage{latexsym}
\usepackage{graphicx}
\usepackage{amsfonts,amssymb}
\usepackage{subfigure}

\begin{document}
\title{\textbf{Scaling Properties in Spatial Networks and its Effects on Topology and Traffic Dynamics}}
\author{Hua Yang, Yuchao Nie, Ying Fan, Yanqing Hu$\footnote{Author for correspondence:yanqing.hu.sc@gmail.com}$, Zengru Di$
\footnote{Author for correspondence:zdi@bnu.edu.cn}$
\\Department of Systems Science, School of management,
\\Center for Complexity Research, Beijing Normal University, Beijing 100875, P.R.China.}
\maketitle

\begin{abstract}
Empirical studies on the spatial structures in several real
transport networks reveal that the distance distribution in these
networks obeys power law. To discuss the influence of the power-law
exponent on the network's structure and function, a spatial network
model is proposed. Based on a regular network and subject to a
limited cost $C$, long range connections are added with power law
distance distribution $P(r)=ar^{-\delta}$. Some basic topological
properties of the network with different $\delta$ are studied. It is
found that the network has the smallest average shortest path when
$\delta=2$. Then a traffic model on this network is investigated. It
is found that the network with $\delta=1.5$ is best for the traffic
process. All of these results give us some deep understandings about
the relationship between spatial structure and network function.

\end{abstract}



\section {Introduction}
In the last few years, the analysis and modeling of networked
systems have received considerable attention within the physics
community, including the World Wide Web, the Internet, and
biological, social, and infrastructure networks\cite{R. Albert,S.N.
Dorogovtsev,M.E.J. Newman}. Some of these networks exist only in an
abstract network where the precise positions of the network nodes
have no particular meaning, such as biochemical networks and social
networks, while many others are different in which nodes have
well-defined positions. In all kinds of network, a particular class
is the spatial network embedded in the real space. Many networks
belong to this class like the neural network\cite{O Sporns},
communication networks\cite{V. Latora}, electric power
grid\cite{G.L. Nakarado}, transportation systems ranging from
airport\cite{R. Guimer¨¤}, street\cite{P. Crucitti} , railway and
subway\cite{M. Marchiori} networks. Most of the previous works in
the studies of complex network have focused on the characterization
of the topological properties or other issues, while the spatial
aspect has received more attention recently \cite{M.T.
Gastner,Alessio Cardillo,Kosmas Kosmidis,Yukio Hayashi}.

Actually, the geography matters greatly. The geography information
of the nodes and the distance between nodes would determine the
characteristics of network and play an important role in the
dynamics process happened on network more or less. Ignoring it would
miss some of these systems' interesting features. Empirical studies
have revealed some interesting phenomena about the spatial structure
of networks. One is the distance distribution of the links obeys
power law. Examples include the Internet\cite{S-H.Yook}, social
communication network\cite{Renaud Lambiotte} and online social
network\cite{Liben}. Recent research on circuit placement showed
that the wire length of real circuits exhibits a power law
distribution\cite{Dambre}. The anatomical distance distributions in
human brain networks can also be well fitted by an exponentially
truncated power-law\cite{He}. More evidence comes from the empirical
research on transportation networks. For Japanese airline
networks\cite{Hayashi}, even there is an exponential decay in
domestic flights, the distance distribution follows power-law when
international flights are added. For the U.S. intercity passenger
air transportation network\cite{Zengwang Xu}, the distribution of
link distance has a power-law tail with exponent is -2.2.

\begin{figure}[!htb]
\center \mbox{
     \subfigure[]
        {\includegraphics[width=5.9cm]{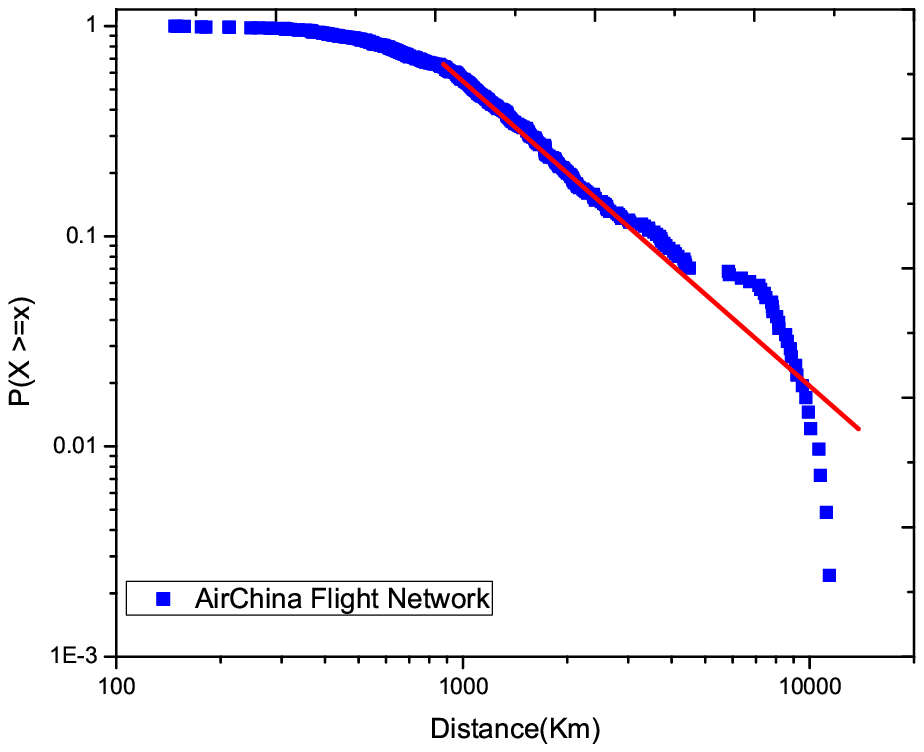}
        \label{1:a}}
        \subfigure[]
        {\includegraphics[width=6cm]{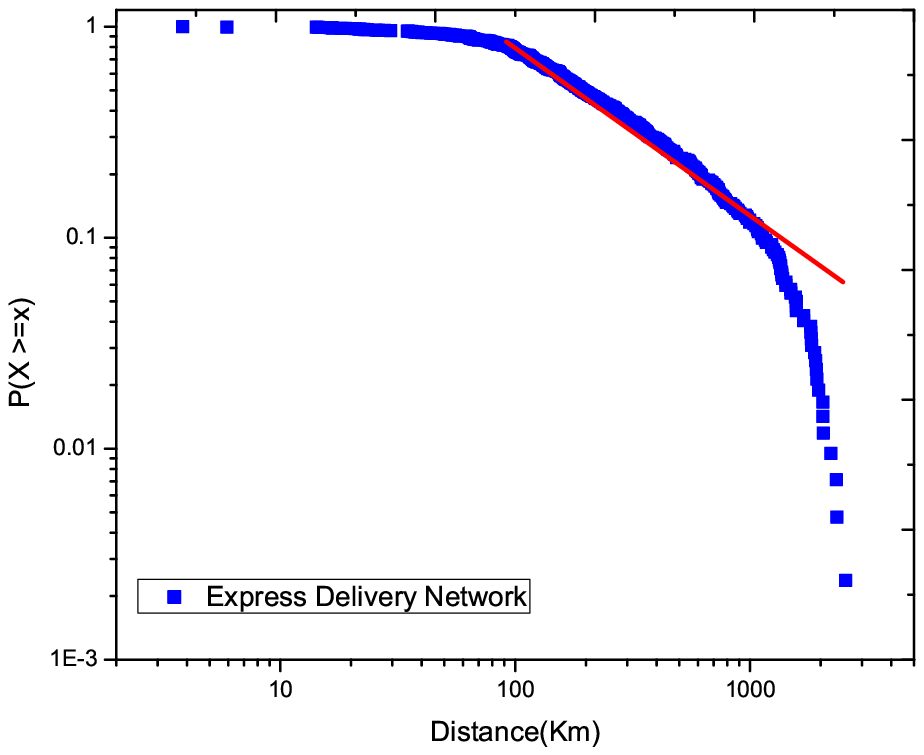}
        \label{1:b}}
       }
\caption{The edge length distribution of (a)the Air China airline
network and (b)the express network}\label{1}
\end{figure}

We also get the data of airline networks for Air China (CA) from its
web site (http://www.airchina.com.cn). The network consists of 177
airports and 411 links including international flights. The distance
distribution shows power law with exponent -2.38 (As shown in
Fig.\ref{1} (a)). Another example is the express delivery network
from a Chinese logistics company. It contains 301 cities and 421
undirected routes managed by the company. We examine the network in
which the nodes are cities with more than one express services
sector and the edges are the delivery route between cities. The
networks' lengths also show power-law distribution with exponents
-1.87 (As shown in Fig.\ref{1} (b)).

Why does the geographical embedded network possess this special
spatial structure? How does the distance distribution affect the
network's function? All these problems are interesting. Analyzing
these problems will help us understand the real spatial network
deeply and benefit us for the design of transport system. The clue
to answer these questions may rely on the consideration of costs and
efficiency.

In spatial embedded networks, especially transport networks, the
connection between nodes are restricted by cost constraints,
reflected through the distance distribution. The cost of
establishing long-range connections between distant spots is usually
higher than the cost of establishing short-range connections. For
electric power grids, the connection cost between farther spots is
even higher, given that in long high-voltage lines a large amount of
energy is lost during the transmission\cite{R Xulvi-Brunet}. So we
can easily understand that the number of short-range connections is
much more than the long-range connections in these networks.

To demonstrate how the distance distribution of the connections
affect the structure, function and the traffic dynamics process of
the networks, we proposed a spatial network model in this paper. The
model takes both the power law distribution of distance and the
total cost of links into account. We construct spatially constrained
networks embedded in geographical space, the distance distribution
of the network obeys power law distribution $P(r)\propto
r^{-\delta}$ and the network has a limited total cost $C$ to create
links. We analysis the spatial network in detail, It shows that the
network has the smallest average shortest path when $\delta=2$ and
it isn't influenced by the value of the total cost $C$ and the size
of networks. Then the traffic model on the network is investigated.
It is found that the network with $\delta=1.5$ is best for the
traffic process. All of these results may explain why some of
spatial networks' exponent is close to 2, and the exponent of
express delivery network is smaller than airline networks.

\section{Spatial network model with limited total cost}

Generally, the cost and efficiency are equally important in
transport networks. The network structure is the result caused by
the tradeoff between cost and efficiency.

The model network is embedded in a $k$-dimensional regular network.
The long range connections is generated from a power law distance
distribution by the approach suggested in\cite{Kosmas Kosmidis}.
Different from previous model\cite{Kosmas Kosmidis}, we introduce a
total cost $C$ to this network model. Every link has a cost $c$
which is linear proportion to its distance $r$. For simplification
and without lose any generalization, the edge cost $c_{i}$ is
represented by its length $r_{i}$ in the model. The network is
constructed under a certain limited total cost $C$. Then $L$ long
range links with power law distance distribution are created under
the condition $\sum^{L}_{i=1}r_{i}=C$. The network is constructed as
following.

\begin{enumerate}
\item $N$ nodes are arranged in a $k$-dimensional lattice. Every node is connected with its nearest neighbors which
can keep every node reachable. And, between any pair of nodes there
is a well defined Euclidean distance.
\item A node $i$ is chosen randomly, and a certain distance
$r (2\leq r\leq N_{max}$, $N_{max}$ is the largest distance between
any nodes in the initial network.) from node $i$ is generated with
probability $P(r)=ar^{-\delta}$, where $a$ is determined from the
normalization condition $\sum^{N_{max}}_{r=2}P(r)=1$.
\item One of the $N_{r}$ nodes that are at distance
$r$ from node $i$ is picked randomly, for example node $j$. An edge
with distance weight $r$ between nodes $i$ and $j$ is created if
there exist no edge between them. We make the multiple connections
between two nodes excluded in this model.
\item  Once the edge is created, a certain cost $c$ is generated. Repeat step 2 and 3 until the total
cost reaches $C$.
\end{enumerate}

\begin{figure}[!htb]
\center \mbox{ \subfigure[]
        {\includegraphics[width=6cm]{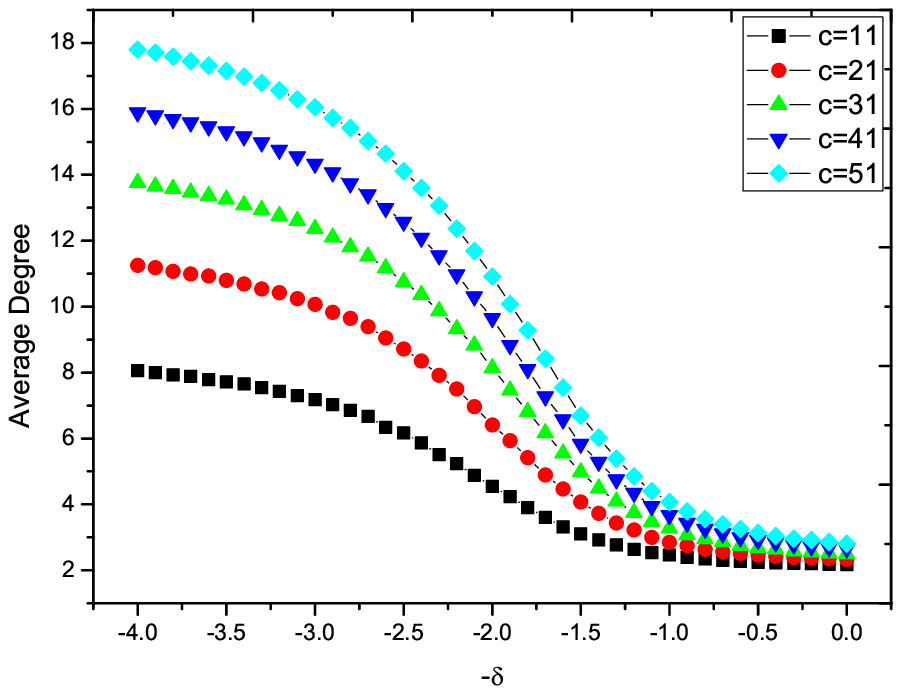}
        \label{2:a}}
        \subfigure[]
        {\includegraphics[width=6cm]{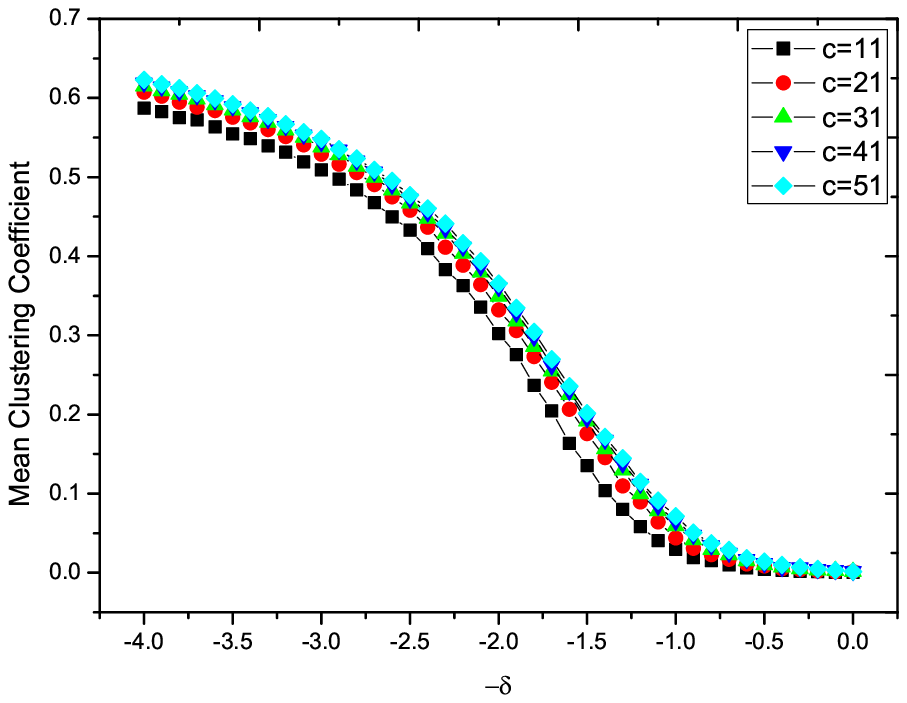}
        \label{2:b}}
        }
\mbox{
        \subfigure[]
        {\includegraphics[width=8cm]{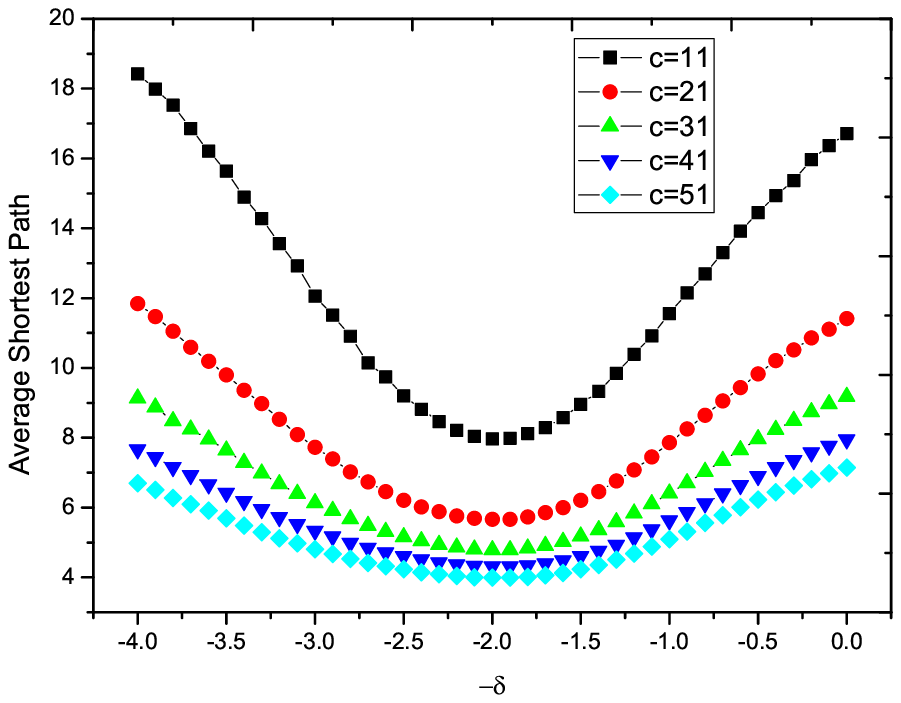}
        \label{2:d}}
        }
\caption{Basic topological properties in the model network. (a),
(b), (c) represent the average degree, average clustering
coefficient, and average shortest path on different total cost
$C=500c (c=11, 21, 31, 41, 51)$ respectively. (c) shows that the
average shortest path reaches minimum when $\delta=2$ with different
total cost $C$.}\label{2}
\end{figure}

In this model, the distance distribution and the total cost play
important roles in the formation of the network. We first focus on
how the topological properties are affect by the two factors. We are
interested in how the power-law exponent $\delta$ influences the
topological properties in the resulted binary network, including the
average degree, mean clustering coefficient, and the average
shortest path of the network. We have simulated the model both in
1-dimensional chain and 2-dimensional lattice, with periodic
boundary condition respectively. They all give same qualitative
results. So in the following, we only report the results of
1-dimensional chain with periodic boundary condition. The network
size is typically $N=500$ and the results are the average of 1000
realizations. The total cost is set as $C=Nc (C=500c)$, where $c$ is
the average cost per node and is set as 11, 21, 31, 41, 51
successively in the simulation. The results are shown in
Figure.\ref{2}. It shows that in the binary networks, the average
degree and mean clustering coefficient decrease with the $\delta$.
Interestingly, the average shortest path reaches its minimum when
$\delta=2$. So when the total cost is limited, the network with
$\delta=2$ has the lowest average shortest path, which may explain
why the power-law exponent of distance distribution is close to 2 in
many airline networks. As we know, in the public transport networks,
travelers prefer less transfer when traveling. The transport network
with $\delta=2$ has the lowest average shortest path in topological
structure, which can make travelers have the least transfer and more
convenience.

We also investigate how the total cost affects the networks
properties. As we know, when the total cost reaches a certain value,
the network with given size would become nearly full connected and
the power law distance distribution will be destroyed. But in fact,
the total cost $C$ is controlled by the average cost per node $c$
and it is generally limited, here we only focus on the effects of
total cost in a reasonable range. When the power-law exponent be
close to -4, in 2 or 3 dimensional space, the underlying network can
not provide enough short long-range connections. It may result in
drop-head phenomenon in the distance distribution. In this model,
most of the long-range connections shows strict power law with
drop-head in a small number of short long-range connections, which
is prevalent in the real world. With the increasing of total cost,
more long rang links are created. The average degree and mean
clustering coefficient increase, while the average shortest path
decreases. One interesting result is that for a certain range of
total cost, the relationship between topological properties and the
parameter $\delta$ keeps the same qualitatively as shown in
Fig.\ref{2}. Then, how this structure is related with the traffic
dynamics? Next section will show us some results.

\section{Traffic process on the model spatial network}
From the analysis of the spatial network above, the average shortest
path in the model network may explain why some public networks's
exponent close to 2 to a certain extent. What will happen when we
consider the traffic dynamics on the above spatial embedded
networks? To investigate the traffic dynamics may be helpful to
understand that the express delivery network has a relatively
smaller exponent in the distance distribution. The express delivery
network is constructed based on all kinds of public transport
networks, especially the airline networks. But it has its own
feature, which is different from the public transport networks.
First, the express delivery network is constructed from the
perspective of overall optimization, while the public transport
networks are constructed by self-organization, based on a local
optimization process\cite{Marc}. Second, the express delivery
network cares really the traffic process on the network. Both
efficiency and cost are important factors to shape the network
structure. Third, same as the traffic model on networks, the
bottleneck of the express network lies in the node, the capability
of the node determine the whole network's efficiency. So to
investigate the traffic flow on such kind of network may help us to
find how the exponent $\delta$ influences the traffic flow on the
model network.

We employ a typical traffic dynamics\cite{Liang Zhao} on this
spatial network. Firstly, generate the underlying network
infrastructure with the method we propose in last section. We also
take the network as an binary one. Then a traffic dynamics is
modeled on the network. All the nodes embedded on the spatial
network are treated as both hosts and routers. We assume that every
node can deliver at most $D$ packets one step toward their
destinations. At each time step, there are $R$ packets generated
homogeneously on the nodes in the system. The packets are delivered
from their own origin nodes to destination nodes by special routing
strategy. There are many kinds of different routing strategies, such
as the shortest pathways routing strategy, the local dynamical
strategy\cite{Mao-Bin Hu} and so on. Here the shortest pathways
routing strategy is used. A packet, upon reaching its destination,
is removed from the system. The order parameter
\begin {equation}
\eta(R)=\lim_{t\rightarrow\infty}\frac{1}{R}\frac{<\Delta
N_{p}>}{\Delta t}
\end {equation}
is used to characterize the phase transition. Here $\Delta
N_{p}=N_{p}(t+\Delta t)-N_{p}(t)$, $<...>$ denotes taking the
average over a time window of width $\Delta t$. $N_{p}(t)$ is the
number of packets in the system at time $t$. We are most interested
in the critical value $R_{c}$ (as measured by the number of packets
created within the network per unit time), where a phase transition
takes place from free flow to congested traffic. This critical value
can best reflect the maximum capability of a system handling its
traffic. In particular, for $R<R_{c}$, the numbers of created and
delivered packets are balanced, leading to a steady free traffic
flow. For $R>R_{c}$, traffic congestion occurs as the number of
accumulated packets increases with time, simply because the
capacities of the nodes for delivering packets are limited.

For simplicity, here we construct a network with 500 nodes under the
total cost $C=500\times11$ and $C=500\times51$, and set every node
has the same delivery ability $D=1$. We will adjust the network
parameters $\delta$ to generate different networks, and then
investigate how $\delta$ affects the critical value $R_{c}$ and the
order parameter $\eta(R)$.

\begin{figure}[!htb]
\center \mbox{ \subfigure[]
        {\includegraphics[width=7cm]{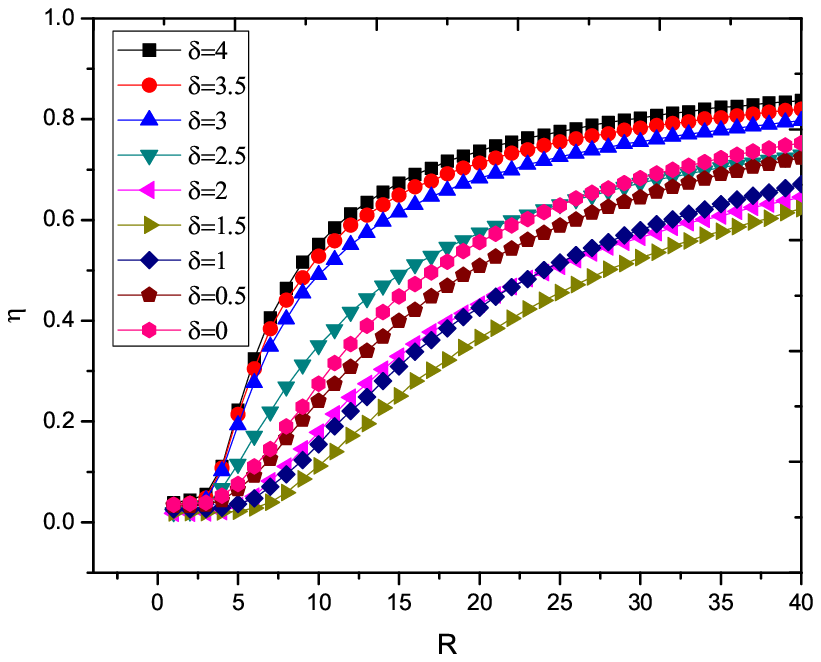}
        \label{3:a}}
     }
\mbox{ \subfigure[]
        {\includegraphics[width=7cm]{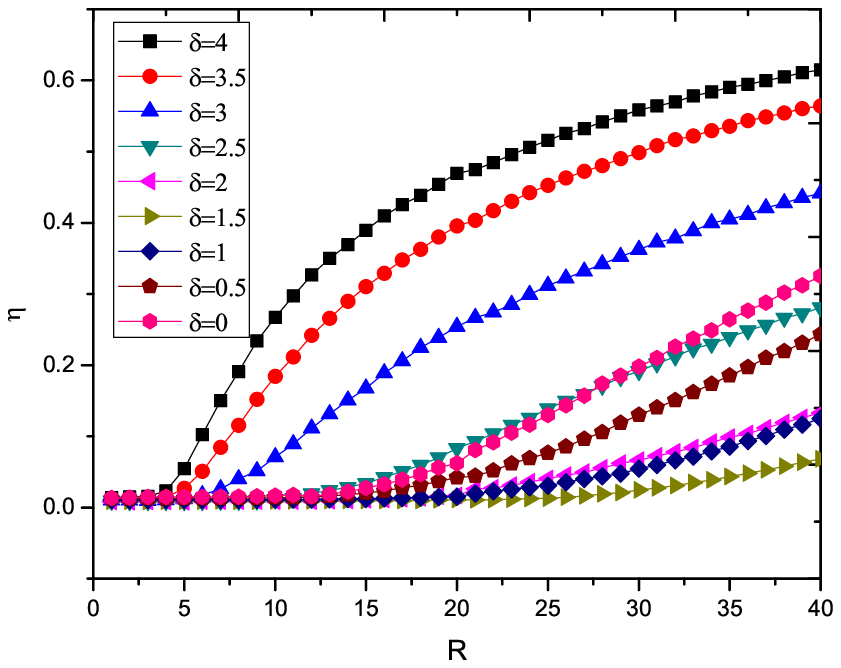}
        \label{3:b}}
     }
\caption{The simulation for the critical value $\eta(R)$ as $R$,
different color mean for different $\delta$. The total cost C is
500$\times$11 in (a) and 500$\times$51 in (b). It shows that with
different total cost, the network with $\delta=1.5$ is best for the
traffic process.}\label{3}
\end{figure}

The simulation results for the critical value $\eta(R)$ as a
function of $R$ on the model networks are reported in the
Figure.\ref{3}. It shows that the network with $\delta=1.5$ has a
biggest $R_{c}$ and the smallest $\eta(R)$.

In previous works, traffic handling capacity of a particular network
has been estimated by a simple analysis method\cite{Liang Zhao}. The
betweenness coefficient of node $v$ can be calculated as
$g(v)=\Sigma_{s\neq t}\frac{\sigma_{st}(v)}{\sigma_{st}}$. Here
$\sigma_{st}$ is the number of shortest paths going from $s$ to $t$
and $\sigma_{st}(v)$ is the number of shortest paths going from $s$
to $t$ and passing through $v$. Note that with the increasing of
parameter $R$ (number of packets generated every step), the system
undergoes a continuous phase transition to a congested phase. Below
the critical value $R_{c}$, there is no accumulation at any node in
the network and the number of packets that arrive at node $u$ is
$Rg_{u}/N(N-1)$ on average. Therefore, a particular node will
collapse when $Rg_{u}/N(N-1)>D_{u}$ where $g_{u}$ is the betweenness
coefficient and $D_{u}$ is the transferring capacity of node $u$.
So, congestion occurs at the node with the largest betweenness. Thus
$R_{c}$ can be estimated as $R_{c}=D_{u}N(N-1)/g_{max}$, where
$g_{max}$ is the largest betweenness coefficient of the network.

\begin{figure}[!htb]
\center \mbox{ \subfigure[]
        {\includegraphics[width=7cm]{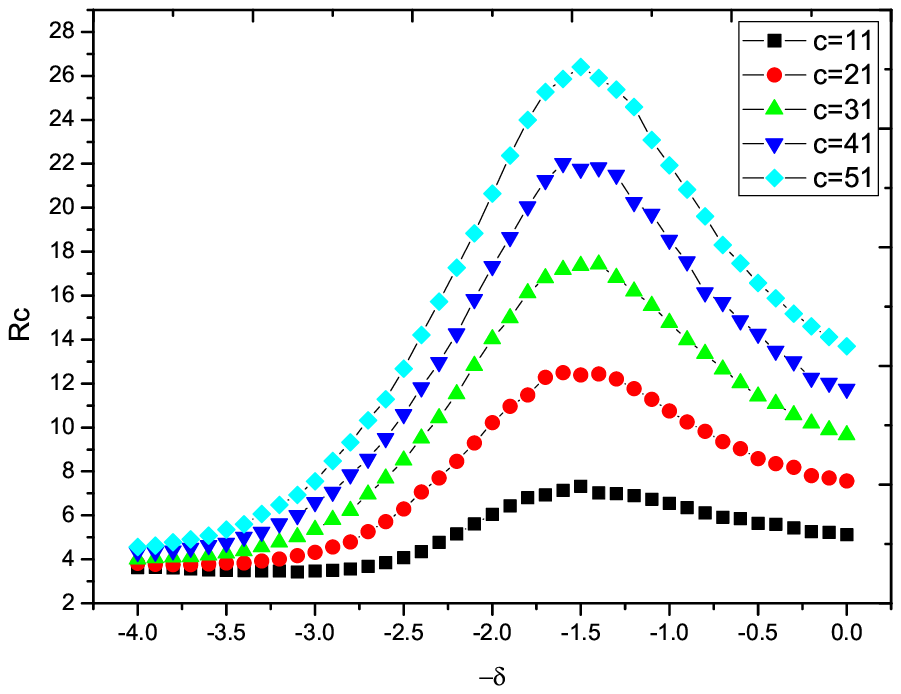}
        \label{4:a}}
     }
\mbox{ \subfigure[]
        {\includegraphics[width=7.1cm]{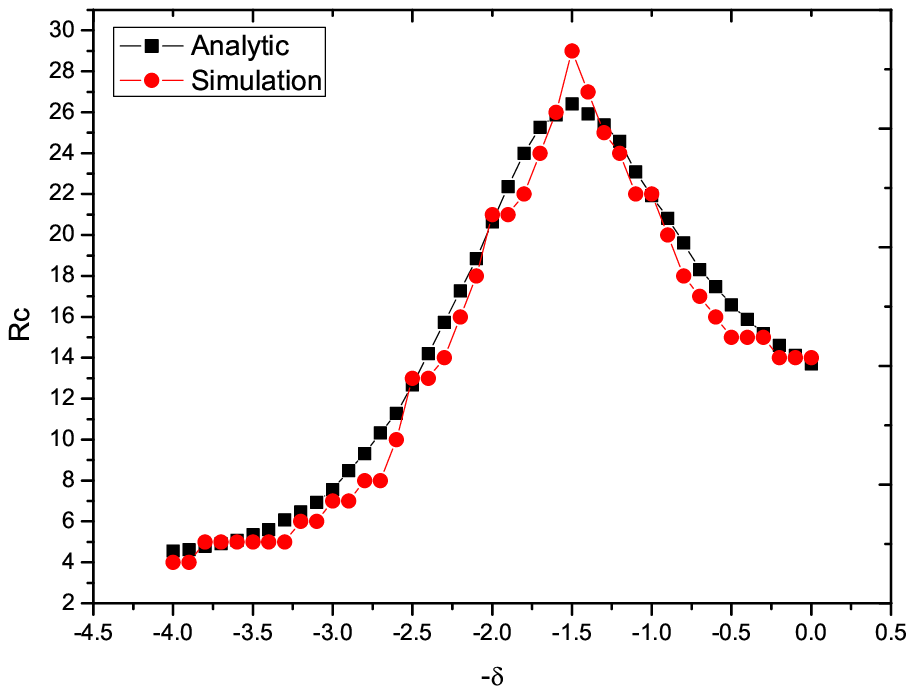}
        \label{4:b}}
     }
\caption{(a) shows the analytical results of $R_{c}$ with parameters
$\delta$. The total cost $C$ increases from bottom-up. (b) shows the
critical $R_{c}$ vs $\delta$ with the total cost $C=500\times51$,
both simulation and analysis indicate that the maximum $R_{c}$
corresponds to $\delta$=1.5.}\label{4}
\end{figure}

In Fig.\ref{4} (a), the analysis results of $R_{c}$ with network
parameters $\delta$ are shown. It indicates that this kind of
network has the biggest $R_{c}$ when $\delta$ is close to 1.5, which
is in good agreement with the simulation results as shown in
Fig.\ref{4} (b).

\section{Conclusions}
In this paper, based on the empirical results about the power law
distance distribution of some spatial networks, a spatial network
model is studied. In a regular lattice, long range connections are
added with the power law distance distribution $P(r)=ar^{-\delta}$
and a limited total cost. Some basic topological properties of the
network generated by the model are investigated. It is found that
the network has the smallest average topological shortest path when
$\delta=2$. This may be the reason of the distance distribution in
airport network shows power law with exponent close to 2, because
people care more about the convenience and prefer less transfer when
traveling by air. Then a traffic model is studied on the model
network. We find that the network with $\delta=1.5$ is most
conducive to the traffic process, although it has not the smallest
average topological shortest path. This may explain why the power
law exponent of the distance distribution in express transportation
network is 1.87. Our results indicate that spatial constraints have
an important influence on the transport networks and should be taken
into account when modeling real complex systems.

\section*{Acknowledgement}
This work is partially supported by 985 Projet and NSFC under the
grant No. $70771011$ and No. $60534080$.

\end{document}